  \documentclass[final,authoryear,5pt, twocolumn]{elsarticle}
  
  \usepackage{amsfonts,amssymb,mathtools,bm,amsthm}
  \usepackage[OT1]{fontenc}
  \usepackage[inline]{enumitem}
  \usepackage{acronym}
  \usepackage{float}
  \usepackage[dvipsnames]{xcolor}
  \usepackage{siunitx, fullpage}
  \usepackage[a4paper=true,breaklinks=true,colorlinks=true,pdfauthor={First Author et al.},pdftitle={Template for manuscripts in Advances in Space Research}]{hyperref} 
  \usepackage[round]{natbib}
  \usepackage{epsfig}

  \newcolumntype{C}[1]{>{\centering}m{#1}}

  \def\eqns#1{\begin{equation*}#1\end{equation*}}
  \def\eqnl#1#2{\begin{equation}\label{#1}#2\end{equation}}
  
  \def\eqnla#1#2{\begin{subequations}\label{#1}\begin{align}#2\end{align}\end{subequations}}
  \ifx\draft\undefined
    \def\eqnsm#1{\begin{multline*}#1\end{multline*}}
    \def\eqnsml#1#2{\begin{multline}\label{#1}#2\end{multline}}
  \else
    \def\eqnsm#1{\begin{equation*}#1\end{equation*}}
    \def\eqnsml#1#2{\begin{equation}\label{#1}#2\end{equation}}
  \fi
  \def\nosmcrun{20~} 
  \ifx\draft\undefined
    \def\imagewidth{0.45\textwidth}
  \else
    \def\imagewidth{0.60\textwidth}
  \fi
  
  \def\one{\mathbf{1}}
  \def\bsa{\bm{a}}

  \def\bsh{\bm{h}}

  \def\bsomega{\bm{\omega}}
  \def\bsp{\bm{p}}

  \def\boS{\mathbf{S}}

  \def\boX{\mathbf{X}}
  \def\bsv{\bm{v}}
  \def\boY{\mathbf{Y}}

  \def\calN{\mathcal{N}}

  \def\bbN{\mathbb{N}}
  \def\bbP{\mathbb{P}}
  \def\bbR{\mathbb{R}}

  \def\ang{\mathrm{ang.}}
  \def\c{\mathrm{c}}
  
  \def\d{\mathrm{d}}
  \def\e{\mathrm{e}}
  \def\eci{\mathrm{eci}}
  \def\eff{\mathrm{eff.}}
  \def\en{\mathrm{en.}}
  \def\i{\mathrm{i}}
  \def\map{\mathrm{map}}
  \def\mod{\mathrm{mod.}}
  \def\n{\mathrm{n}}
  
  \def\orb{\mathrm{orb.}}
  \def\r{\mathrm{r}}
  \def\ric{\mathrm{ric}}

  \def\tle{\mathrm{tle}}
  \def\tr{\mathrm{t}}
  \def\rad{\mathrm{rad.}}

  \def\given{\,|\,}

  \graphicspath{{Figures/}}
  \restylefloat{figure} 
  
  \acrodef{amr}[AMR]{Area-to-Mass Ratio}
  \acrodef{eci}[ECI]{Earth-Centered Inertial}
  \acrodef{fov}[FoV]{Field of View}
  \acrodef{gps}[GPS]{Global Positioning System}
  \acrodef{jspoc}[JSpOC]{Joint Space Operations Center}
  \acrodef{leo}[LEO]{Low-Earth Orbit}
  \acrodef{map}[MAP]{Maximum A Posteriori}
  \acrodef{mht}[MHT]{Multiple Hypothesis Tracking}
  \acrodef{opm}[o.p.m.\@]{outer probability measure}
  \acrodef{pdf}[pdf]{probability density function}
  \acrodef{resp}[resp.]{respectively}
  \acrodef{ric}[RIC]{Radial-Intrack-Crosstrack}
  \acrodef{rso}[RSO]{Resident Space Object}
  \acrodef{smc}[SMC]{Sequential Monte Carlo}
  \acrodef{ssa}[SSA]{Space Situational Awareness}
  \acrodef{tle}[TLE]{Two-Line Element}
  \acrodefplural{tle}[TLEs]{Two-Line Elements}
  \acrodef{usscom}[\textsc{USStratCom}]{U.S. Strategic Command}
  \acrodef{wrt}[w.r.t.]{with respect to}
  
  \journal{Advances in Space Research}
  
\begin{document}
  
  \begin{frontmatter}
  
    \title{Physics and Human-Based Information Fusion for Improved Resident Space Object Tracking}

    \author[manu]{Emmanuel Delande\corref{cor}}
    \address[manu]{Institute for Computational Engineering and Sciences, The University of Texas at Austin, U.S.A.}
    \cortext[cor]{Corresponding author}
    \ead{edelande@ices.utexas.edu}

    \author[jerem]{J\'{e}r\'{e}mie Houssineau}
    \address[jerem]{Department of Statistics and Applied Probability, National University of Singapore, Singapore.}
    \ead{stahje@nus.edu.sg}

    \author[moriba]{Moriba Jah}
    \address[moriba]{Department of Aerospace Engineering and Engineering Mechanics, The University of Texas at Austin, U.S.A.}
    \ead{moriba@utexas.edu}

    \begin{abstract}
      Maintaining a catalog of \acp{rso} can be cast in a typical Bayesian multi-object estimation problem, where the various sources of uncertainty in the problem -- the orbital mechanics, the kinematic states of the identified objects, the data sources, etc. -- are modeled as random variables with associated probability distributions. In the context of \acl{ssa}, however, the information available to a space analyst on many uncertain components is scarce, preventing their appropriate modeling with a random variable and thus their exploitation in a \ac{rso} tracking algorithm. A typical example are human-based data sources such as \acp{tle}, which are publicly available but lack any statistical description of their accuracy. In this paper, we propose the first exploitation of \emph{uncertain} variables in a \ac{rso} tracking problem, allowing for a representation of the uncertain components reflecting the information available to the space analyst, however scarce, and nothing more. In particular, we show that a human-based data source and a physics-based data source can be embedded in a unified and rigorous Bayesian estimator in order to track a \ac{rso}. We illustrate this concept on a scenario where real \acp{tle} queried from the \acl{usscom} are fused with realistically simulated radar observations in order to track a \acl{leo} satellite.
    \end{abstract}

  \end{frontmatter}

  \parindent=0.5 cm
  \acresetall
  
  \section{Introduction} \label{sec:intro}
    \ac{ssa} concerns itself with having the actionable knowledge to predict, deter, avoid, operate through, recover from, or attribute cause to the loss, disruption, degradation, or denial of space services, capabilities, or activities. Maintaining a catalog of so-called \acp{rso} can be cast as a typical \emph{multi-object detection and tracking problem}. The individual \emph{state} of each object consists of kinematic and relevant model parameters to the construction of the catalog and/or its exploitation for various activities related to space, such as mission planning, conjunction analysis to prevent collisions, etc. Typically, the individual state of an orbiting object describes a kinematic relationship (Kepler elements, or position/velocity coordinates), but may include additional characteristics such as its attitude, its ballistic coefficient, etc.
    
    The main challenge to the resolution of most detection and tracking problems is to appropriately address the various sources of \emph{uncertainty} involved in its different components, due to the limited knowledge in the signal processing chain affecting the acquisition of information on the population of objects. The \ac{ssa} problem is no different, and the relevant sources of uncertainty can be sorted in two broad categories:
    \begin{enumerate}[wide]
      \item To first order, the perturbing accelerations on the \ac{rso} population are known, gravity being the dominant perturbing source. However, no orbital propagator is able to produce the exact dynamical state of an \ac{rso}, since the various physical perturbations affecting orbiting trajectories are either approximated, discarded on purpose for the sake of algorithmic efficiency, or simply unknown to the space analyst. In addition, satellites owned by others may transition to new orbital paths unknown to the analyst. New \acp{rso} entering some specific orbital region, either through a collision or a new launching event, are not all accounted for and thus the number of \acp{rso} remain uncertain. 
      \item The mechanisms through which observations (or opinions) are produced by the data sources are partially known, at best. Traditional, physics-based sensors like radars, telescopes, or cameras are subject to missed detections, false positives, observation noise, from which some statistical description is usually available. On the other hand, the \acp{tle} produced by the \ac{usscom} are publicly available, but no information regarding the accuracy of the orbital elements or the confidence in the object labeling are available.
    \end{enumerate}
    The Bayesian estimation framework is a very popular approach to solve multi-object detection and tracking problems, and the origin of the overwhelming majority of modern tracking algorithms. Its key features are that
    \begin{enumerate}[label=\alph*), wide]
      \item the uncertainty on each component of the system is modeled with a \emph{random variable} and characterized with a \emph{probability distribution}, and
      \item a probability distribution can be updated sequentially, or \emph{filtered}, with the availability of new information regarding the corresponding random variable.  
    \end{enumerate}
    Bayesian filters then maintain a probabilistic description of the population of objects and propagate it through time, updating it whenever new observations are collected from a data source, also described through a probabilistic model. The characteristics of a specific Bayesian filter primarily depends on the nature of the probabilistic description representing the population of objects; most notably, through individual tracks \citep{Blackman_SS_2004}, labeled random finite sets \citep{Vo_BT_2013}, or stochastic populations \citep{Houssineau_J_2016_1, Delande_E_2016_2_1}. Applications of these various multi-object filtering techniques to the context of \ac{ssa} can be found in \cite{DeMars_KJ_2012_3}, in \cite{Jones_BA_2015_2, Jones_BA_2016_1}, and in \cite{Delande_E_2018_1, Delande_E_2017_1_2}, respectively.

    One of the key challenges of Bayesian estimation is to maintain an appropriate representation of the uncertainty on the components of the system, especially when information about their state is scarce. A typical example in the context of \ac{ssa} is the initial orbit determination procedure through which the possible values for the dynamical state (position, velocity coordinates) of a newly-detected \ac{rso} are restricted to a specific subspace known as the admissible region in \cite{DeMars_KJ_2013_2}. The physical considerations behind the initial orbit determination procedure do not provide information on whether any dynamical state within the admissible region is likelier than the other, but it does \emph{not} indicate that all states within the admissible region are equally likely \emph{either}\footnote{It turns out that they are not, since a more restrictive initial orbit determination procedure can lead to a \emph{constrained} admissible region \citep{DeMars_KJ_2012_3}.}. However, a probabilistic interpretation of the \ac{rso}'s initial state can lead to a uniform probability distribution on the admissible region, thus producing a description of the \ac{rso}'s state that is \emph{not} inferred from the information available to the analyst. It thus appears that \acp{pdf} are inappropriate tools to describe admissible regions, as argued in a recent study \citep{Worthy_JLIII_2017_2}. Another example is the exploitation of human-based or semantic data sources such as \acp{tle} or natural languages statements: they could provide a wealth of information regarding \acp{rso}, but the lack of statistical information on their accuracy/truthfulness makes their probabilistic representation, and thus their integration to a Bayesian tracking filter, difficult and largely unexplored to this day.
    
    Alternatives to the standard probabilistic representation of uncertainty exist, such as fuzzy logic, imprecise probabilities, possibility theory, fuzzy random sets, and Dempster-Shafer theory \citep{Zadeh_LA_1965_1, Walley_P_1991_1, Dempster_AP_1967_1, Shafer_G_1976_1, Dubois_D_1983_1, Yen_J_1990_1, Friedman_N_2001_1}; recently, the Dempster-Shafer theory has been exploited to approach the initial orbit determination problem with admissible regions in \cite{Worthy_JLIII_2017_1}. While these methods have been widely used to describe the uncertainty referring to a \emph{fixed} unknown state, their exploitation to the estimation of \emph{dynamical} systems is less straightforward, limiting their applicability to sequential estimation problems -- that is, to the design of detection and tracking filters.

    A recent alternative is given by the \emph{\acp{opm}} as introduced in \cite{Houssineau_J_2017_1}. They aim at proposing a ``prejudice-free'' representation of the uncertainty on a system through a natural construction that is derived from the available information, and nothing more. Built from fundamental tools of measure theory, \acp{opm} are compatible with the Bayesian filtering framework and can be integrated in a complex system where some uncertain components are described with \acp{pdf}, while others are described with \acp{opm}.
    
    In this paper, we will focus on the representation of uncertainty for \ac{ssa} data sources supporting a \ac{rso} tracking algorithm. More specifically, we will show that radar observations (physics-based information) and \acp{tle} (human-based information) can both be represented by \acp{opm} and integrated into a Bayesian filtering algorithm, where the information on the \ac{rso}'s state is maintained with a probability distribution. Sec.~\ref{sec:uncertain} presents the concepts of \ac{opm} and possibility function, and Sec.~\ref{sec:filtering} describes the filtering equations of the single-object Bayesian tracking algorithm exploiting possibility functions. Sec.~\ref{sec:modeling} focuses on the modeling of the data sources relevant to this paper, i.e., a radar with Doppler effect and a ``\ac{tle}-generator''. Then, Sec.~\ref{sec:scenario} describes the construction of a target tracking scenario with realistically-simulated radar observations and real \ac{tle} data provided by the \ac{usscom}, and gives a detailed implementation of the single-target tracking algorithm. Sec.~\ref{sec:simulation} presents the filtering results. A discussion on future works follows in Sec.~\ref{sec:discussion}, and Sec.~\ref{sec:conclusion} concludes.
    
  \section{Uncertain variables and \acp{opm}} \label{sec:uncertain}
    \subsection{Random and non-random uncertainty}
      Assume some system, whose state is described by some state space $\boX$, on which some operator -- say, a space analyst -- possesses some knowledge. We shall decompose the uncertainty on the system's state in two different parts:
      \begin{itemize}[wide]
	\item The \emph{random} uncertainty is the inherent component, that exists as a natural disposition of the system and independently of the analyst, sometimes known as the \emph{aleatoric} uncertainty,
	\item The \emph{non-random} uncertainty is the extraneous component, that exists only \ac{wrt} the information possessed by the analyst, sometimes known as the \emph{epistemic} uncertainty.
      \end{itemize}
      The random component can be seen as the remaining uncertainty on a system's state, once the analyst possesses all the information on the system that could be acquired. It is unclear if physical systems described on the macroscopic level, such as an \ac{rso}'s dynamical state, can be anything but deterministic, or if there are random effects that cannot be explained assuming perfect knowledge of the physical laws governing the orbital motion model. It is, in the end, irrelevant to the estimation problem that focuses on reducing the non-random -- or epistemic -- uncertainty.
    
    \subsection{Possibility functions}
\begin{figure}[ht]
	\centering
	\includegraphics[width=\imagewidth]{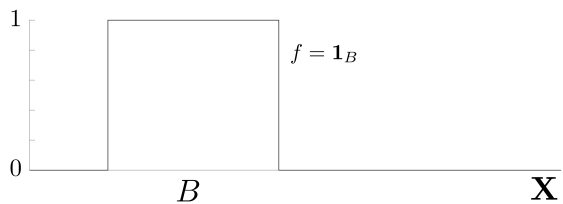}
	\caption{An illustration of a an upper bound $f$. The analyst infers that the object's state lies within some region $B$, but nothing more.\label{fig:poss_1}}	
      \end{figure}

      The random uncertainty, inherent to the system, shall be \emph{characterized} by a probability distribution $p$ on the state space $\boX$.\footnote{For the sake of simplicity, we shall assume that $p$ admits a density \ac{wrt} some reference measure on $\boX$, also called $p$, and we shall handle \acp{pdf} rather than probability distributions, whenever applicable, in the rest of the paper.}
      The non-random uncertainty reflects the information on the system's state possessed by the analyst. In the simplest case, it is maintained by a single function $f$ satisfying 
      \eqnl{eq:upper_bound}
      {
	\int_B p(x) \d x \leq \max_{x \in B} f(x),
      }
      for any appropriate\footnote{The subset must be measurable, for the integral to be well-defined. For the sake of simplicity this requirement on subsets, whenever applicable, will be omitted in the rest of the paper.} subset $B$ of $\boX$. The function $f$ is assumed to have maximum $1$ for the sake of convenience. In other words, while the \ac{pdf} $p$ describing the system's state is inaccessible to the analyst, they maintain an \emph{upper bound} $f$ reflecting the information they possess about the system. A simple example of upper bound is given in Fig.~\ref{fig:poss_1}.
      
      The non-negative, measurable functions on $\boX$ with maximum $1$ are also called \emph{possibility functions} \citep{Houssineau_J_2015_1_2}, or possibility distribution as in \cite{Dubois_D_1983_1}. Box-shaped possibility functions such as illustrated in Fig.~\ref{fig:poss_1} are convenient to represent negative evidence of the system's state, i.e., to exclude regions of the state space as possible values without inferring anything else -- most notably, \emph{without} assuming that all the remaining values are equally probable.
      
\begin{figure}[ht]
	\centering
	\includegraphics[width=\imagewidth]{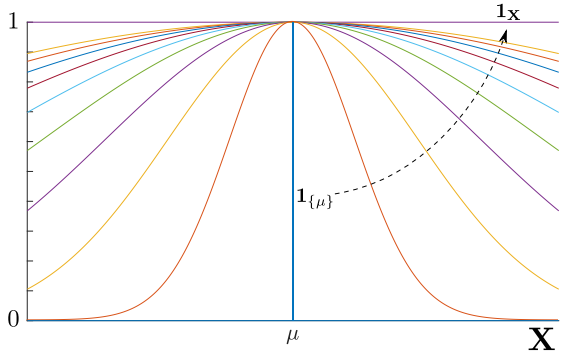}
	\caption{An illustration of a family of Gaussian possibilities $f_n(\cdot) = \bar\calN(\cdot; \mu, S_n)$, with increasing covariance $S_n$. When $S_n\rightarrow 0$, $f_n$ converges (point-wise) to the indicator function $\one_{\{\mu\}}$; when $S_n\rightarrow \infty$, $f_n$ converges (uniformly) to the indicator function $\one_{\boX}$. \label{fig:poss_gaussian}}	
      \end{figure}

      In many situations, such as the modeling of a physics-based sensor, more refined possibility functions are necessary. Similarly as for probability distributions, we define \emph{Gaussian possibility functions} as functions on $\bbR^d$, $d > 0$, of the form
      \eqns{
        \bar\calN(x; \mu, S) = \exp\Big( -\dfrac{1}{2} (x-\mu)^{\tr}S^{-1}(x-\mu)\Big),
      }
      for some $\mu \in \bbR^d$ and some $d\times d$ positive-definite matrix $S$, where $\cdot^{\tr}$ denotes the matrix transposition. For the sake of convenience, we will refer to $\mu$ and $S$ as the mean and variance of $\bar\calN(x; \mu, S)$, although these concepts do not bear the statistical meaning of their namesakes for probability distributions. A family of Gaussian possibility functions is illustrated in Fig.~\ref{fig:poss_gaussian}.
      
      Fig.~\ref{fig:poss_gaussian} also illustrates two particular possibility functions. The indicator function $\one_{\{\mu\}}$ denotes full knowledge about the system, known to have state $\mu$ \emph{almost surely}, since the \ac{pdf} $p$ with mass concentrated in $\mu$ is the \emph{only} \ac{pdf} on $\boX$ satisfying \eqref{eq:upper_bound} when $f = \one_{\{\mu\}}$. Conversely, the indicator $\one_{\boX}$ denotes the total absence of information about the system, since \emph{every} \ac{pdf} $p$ on $\boX$ satisfy \eqref{eq:upper_bound} when $f = \one_{\boX}$. The latter possibility represents a truly uninformative uncertainty, for one does not infer anything regarding the system's state, and in particular does \emph{not} infer that all the states in $\boX$ are equally probable: indeed, the uniform probability distribution on $\boX$ does satisfy \eqref{eq:upper_bound} when $f = \one_{\boX}$, but so do any other probability distribution on $\boX$.
    
   \subsection{Uncertain variables and \acp{opm}}   
      In the general case, the information on some system's state maintained by an analyst can assume a complex form built from a combination of possibility functions. More formally, it can be represented through a probability distribution $P$ on the space of possibility functions such that
      \eqnl{eq:upper_bound_general}
      {
	\int_B p(x) \d x \leq \int \max_{x \in B} f(x) P(\d f),
      }
      for any subset $B$ of $\boX$. In practical problems, a finite combination of possibility functions is often sufficient to describe the uncertainty regarding many systems; in the scope of this paper, we will see in Sec.~\ref{sec:modeling} that a single possibility function is enough to describe various data sources relevant to the context of \ac{ssa}.     
    
      The information given by $P$ induces an \emph{\acl{opm}} $\bar{P}$ on $\boX$ given by
      \eqnl{eq:possibilty_outer_measure}
      {
	\bar{P}(B) = \int \max_{x \in B} f(x) P(\d f),
      }
      for any subset $B$ of $\boX$. The scalar $0 \leq \bar{P}(B) \leq 1$ reads as the \emph{possibility} that the system's state is in $B$. This is to be contrasted with the \emph{probability measure} $\bbP$ \emph{characterizing} the inherent randomness of the object's state, independently of the information possessed by the analyst, and given by
      \eqnl{eq:probability_measure}
      {
	\bbP(X \in B) = \int_B p(x)\d x.
      }
      From \eqref{eq:upper_bound_general} it follows that
      \eqnl{eq:proba_and_poss}
      {
	1 - \bar{P}(\boX \setminus B) \leq \bbP(X \in B) \leq \bar{P}(B),
      }
      for any subset $B$ of $\boX$, where $\cdot\setminus\cdot$ denotes the set difference. In other words, an analyst with information represented by $\bar{P}$ cannot (in the general case) evaluate the probability for the object to lie within $B$, but they may \emph{bound} its value.
           
      Note that an \ac{opm} represents the uncertainty of an analyst in the state of some system that might not be random. In that case, the system has a fully deterministic behavior and the mass of the \ac{pdf} $p$ in \eqref{eq:probability_measure} is concentrated to some state $x \in \boX$ -- unknown to the analyst, as they are unaware of the nature of $p$. For this reason, the notion of \emph{uncertain} variables is introduced as a replacement for \emph{random} variables to describe uncertain systems \citep{Houssineau_J_2017_1}.
      
    \subsection{A simple example of \acp{opm}}
\begin{figure}[ht]
	\centering
	\includegraphics[width=\imagewidth]{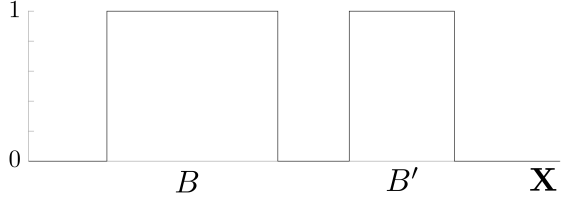}
	\caption{Two distinct levels of information based on the subsets $B$ and $B'$: an analyst may know the probability of the events $X \in B$ and $X \in B'$, or they may only know that $X \in B \cup B'$ \emph{almost surely}.\label{fig:poss_comb}}
      \end{figure}

      We shall describe two distinct levels of information, supported by the same subsets $B$, $B'$ of the state space illustrated in Fig.~\ref{fig:poss_comb}.
      
      Suppose first that the information maintained by the analyst is based solely on the possibility function $\one_{B \cup B'}$ -- that is, the probability measure $P$ in \eqref{eq:possibilty_outer_measure} gives weight $1$ to the singleton $\{\one_{B \cup B'}\}$. Following \eqref{eq:possibilty_outer_measure}, the information is represented by the \ac{opm}
      \eqnl{eq:opm_ex_1}
      {
	\bar{P}_1(A) = [A \cap (B \cup B') \neq \emptyset],
      }
      where $[\cdot]$ is the Iverson bracket\footnote{$[E] = 1$ is the event $E$ is true, and zero otherwise.}. The analyst wishes to gauge the probability that the object's state lies within $B\cup B'$: from Eq.~\eqref{eq:opm_ex_1} we have
      \eqns
      {
	\bar{P}_1(\boX \setminus (B \cup B')) = 0,\quad \bar{P}_1(B \cup B') = 1,
      }
      and from Eq.~\eqref{eq:proba_and_poss} the analyst concludes that
      \eqns
      {
	1 \leq \bbP(X \in B \cup B') \leq 1,
      }
      that is, the object lies in $B\cup B'$ \emph{almost surely}. The analyst then wishes to inquire about the presence of the object in some \emph{smaller} region $C \subsetneq B\cup B'$: from Eq.~\eqref{eq:opm_ex_1} we have
      \eqns
      {
	\bar{P}_1(\boX \setminus C) = 1,\quad \bar{P}_1(C) = 1,
      }
      and from Eq.~\eqref{eq:proba_and_poss} the analyst concludes that
      \eqns
      {
	0 \leq \bbP(X \in C) \leq 1,
      }
      i.e., the analyst is clueless about the distribution of the object \emph{within} $B \cup B'$. In conclusion, the analyst infers that the object is in $B\cup B'$ with certainty, but \emph{nothing more}.
       
      Suppose now that the information maintained by the analyst is based on the two possibility functions $\one_B$, $\one_{B'}$, with respective weights $\alpha$ and $1 - \alpha$, where $0 \leq \alpha \leq 1$. Following \eqref{eq:possibilty_outer_measure}, the information is now represented by the \ac{opm}
      \eqnl{eq:opm_ex_2}
      {
	\bar{P}_2(A) = \alpha[A \cap B \neq \emptyset] + (1 - \alpha)[A \cap B' \neq \emptyset].
      }
      The analyst wishes to gauge the probability that the object's state lies within $B$: from Eq.~\eqref{eq:opm_ex_2} we have
      \eqns
      {
	\bar{P}_2(\boX \setminus B) = 1 - \alpha,\quad \bar{P}_2(B) = \alpha,
      }
      and from Eq.~\eqref{eq:proba_and_poss} the analyst concludes that
      \eqns
      {
	\alpha \leq \bbP(X \in B) \leq \alpha,
      }
      that is, the object lies in $B$ with probability $\alpha$. The analyst then wishes to inquire about the presence of the object in some \emph{smaller} region $C \subsetneq B$: from Eq.~\eqref{eq:opm_ex_2} we have
      \eqns
      {
	\bar{P}_2(\boX \setminus C) = 1,\quad \bar{P}_2(C) = \alpha,
      }
      and from Eq.~\eqref{eq:proba_and_poss} the analyst concludes that
      \eqns
      {
	0 \leq \bbP(X \in C) \leq \alpha.
      }
      i.e., the analyst is clueless about the distribution of the object \emph{within} $B$. A similar reasoning applies in region $B'$; in conclusion, the analyst infers that the object is in $B$ with probability $\alpha$, in $B'$ with probability $1 - \alpha$, but \emph{nothing more}.
    
    \subsection{From \acp{opm} to \acp{pdf}}  
      We see in the example illustrated in the previous section that $\bar{P}_2$ provides a more refined level of information on the system's state than $\bar{P}_1$, because it imposes tighter constraints on the system's underlying \ac{pdf} $p$. In the extreme case where an \ac{opm} $\bar{P}_{\infty}$ imposes constraints on all the singletons of the state space, i.e., when the associated probability measure $P_{\infty}$ in \eqref{eq:possibilty_outer_measure} has support in the possibility functions of the form $\one_{\{x\}}$, where $x \in \boX$, then it can be shown that $P_{\infty}$ induces a \ac{pdf} $p_{\infty}$ on $\boX$ such that
      \eqnl{eq:P_extreme}
      {
	\int_B p_{\infty}(x) \d x = \int \max_{x \in B} f(x) P_{\infty}(\d f) = \bar{P}_{\infty}(B),
      }
      for any subset $B$ of $\boX$. That is, the \ac{opm} $\bar{P}_{\infty}$ is equivalent to the probability measure $\bbP$ in \eqref{eq:probability_measure}, and thus the uncertainty of the analyst in the system's state reduces to the system's inherent random component characterized by $p_{\infty}$. In other words, the analyst has nothing more to learn about the system's state. In particular, if $p_{\infty}$ has its mass concentrated in some point $x^{*} \in \boX$, the analyst concludes that the system is deterministic and has state $x^{*}$ \emph{almost surely}.
      
      This connection between \acp{opm} and \acp{pdf} is of practical importance. It highlights the fact that the distinction between random and non-random uncertainty in some component of a complex system can be purposely ignored for the sake of simplicity; this component is then represented with a usual random variable and described with a level of information that amounts to a probability distribution. As we shall see in Sec.~\ref{sec:filtering}, it allows for the representation of each uncertain component of a detection and tracking problem -- object's state, prediction model, observation model specific to each sensor -- with an uncertain variable, that \emph{may or may not} reduce to a random variable, in a single coherent Bayesian filtering framework.
      
  \section{Bayesian filtering with uncertain variables} \label{sec:filtering}  
    \subsection{Conditional possibility functions and \acp{opm}}
      The formulation of Bayesian inference algorithms rely on the definition of appropriate conditional \acp{pdf} describing the motion and the observation of the object of interest. It is then natural to introduce conditional possibility functions of the form $f( \cdot \given x')$ describing an uncertain variable $X$ given the realization $x'$ in a state space $\boX'$ of another uncertain variable $X'$ and verifying
      \eqns{
	\max_{x \in \boX} f(x \given x') = 1
      }
      for any $x' \in \boX'$. Conditional possibility functions verify the same sort of property as conditional \acp{pdf}: letting $f'$ be a possibility function describing $X'$, it holds that
      \eqnl{eq:poss_pred}
      {
	f(x) = \max_{x' \in \boX'} f(x \given x') f'(x')
      }
      is a possibility function describing the state of $X$ in $\boX$, and that
      \eqnl{eq:poss_up}
      {
	f'(x' \given x) = \dfrac{f(x \given x') f'(x')}{\max_{x' \in \boX'} f(x \given x') f'(x')}
      }
      is a possibility function describing the state of $X'$ given a realization $x \in \boX$ of $X$. Note that Eq.~\eqref{eq:poss_pred} is the analogue of the prediction formula, Eq.~\eqref{eq:poss_up} is the analogue of Bayes' theorem, for possibility functions. Following the approach considered in \cite{Houssineau_J_2018_2}, conditional \acp{opm} can also be defined as
      \eqns{
	\bar{P}( B \given X' = x') = \int \max_{x \in B} f(x \given x') P(\d f \given X')
      }
      where $P(\cdot \given X')$ is a probability measure on the space of conditional possibility functions\footnote{The conditioning on $X'$ indicates the nature of the possibility functions in the support of $P(\cdot \given X')$ but does not make it a function of $x'$.}.
      
    \subsection{Bayesian filtering equations for \acp{opm}}
      Suppose that an analyst studies some object with state in some space $\boX$, described at time $k-1$ (\ac{resp} $k$) by the uncertain variable $X_{k-1}$ (\ac{resp} $X_{k}$). Besides, some sensor with observations\footnote{Since \acp{tle} are not generated by a physics-based sensor, a more accurate term to describe one \ac{tle} point would perhaps be ``opinion''. For the sake of simplicity, the terms ``sensor'' and ``observation'' are usually employed throughout this paper regardless of the nature of the data source.} in some space $\boY_k$ observes the scene, and is described by the uncertain variable $Y_k$.
      
      Suppose that the information that the analyst possesses on the object's state at time $k - 1$ is represented by an \ac{opm} $\bar{P}_{k-1}$, and the information they possesses on the object's evolution between time $k-1$ and $k$ is represented by a conditional \ac{opm} $\bar{M}_{k}(\cdot \given X_{k-1} = x')$, depending on a realization $x'$ of $X_{k-1}$. The predicted information the analyst possesses on the object's state at time $k$ is then described by the \ac{opm}
      \eqnsml{eq:pred_opm}
      {
	\bar{P}_{k\given k-1}(B) =
	\\
	\int \max_{x \in B} \big(\max_{x' \in \boX'}f(x \given x') f'(x') \big) M_{k}(\d f \given X_{k-1}) P_{k-1}(\d f'),
      }
      Suppose that the information that the analyst possesses on the sensor's observation process is represented by a conditional \ac{opm} $\bar{O}_k( \cdot \given X_k = x)$, depending on a realization $x$ of $X_k$, and that an observation $y \in \boY_k$ is collected from the sensor. The updated information the analyst possesses on the object's state at time $k$ is then described by the \ac{opm}
      \eqnsml{eq:upd_opm}
      {
	\bar{P}_{k}(B\given Y_k = y) =
	\\
	\dfrac{\int \max_{x \in B} \big( f(y \given x) f'(x) \big) O_k(\d f \given X_{k}) P_{k\given k-1}(\d f')}{\int \max_{x \in \boX} \big( f(y \given x) f'(x) \big) O_k(\d f \given X_{k}) P_{k\given k-1}(\d f')}.
      }

      The equations \eqref{eq:pred_opm}, \eqref{eq:upd_opm} are the analogues of the time prediction and data update equations for \acp{pdf}, but for \acp{opm}; that is, they are the Bayesian filtering equations for an estimation problem in which the three sources of uncertainty -- the object's state, the object's evolution, the sensor's observation process -- are described with arbitrary uncertain variables, that may or may not reduce to random variables.
      
      Detection and tracking problems in the context of \ac{ssa} present many challenges, whose resolution with Bayesian filters may benefit from approaching these three types of uncertainty with uncertain variables (this is briefly discussed in Sec.~\ref{sec:discussion}). In the context of this paper, however, we shall simply illustrate the benefit from modeling the sensors' observation process with a single possibility function; in particular, both the information on the object's state and the object's evolution will be represented with usual \acp{pdf}.
      
    \subsection{A simpler case of Bayesian filtering equations}        
      We shall suppose, then, that the \ac{opm} $\bar{P}_{k-1}$ on the object's state at time $k-1$ reduces to a \ac{pdf} $p_{k-1}$ on $\boX$, and the \ac{opm} $\bar{M}_{k}(\cdot\given X_{k-1} =  x')$ on the object's evolution reduces to a transition kernel $m_{k}(\cdot|x')$ on $\boX$, depending on a realization $x'$ of $X_{k-1}$. The prediction formula \eqref{eq:pred_opm} then reduces to
      \eqnl{eq:pred_red}
      {
	p_{k\given k-1}(x) = \int m_k(x\given x')p_{k-1}(x')\d x',
      }
      where $p_{k\given k-1}$ is the \ac{pdf} describing the predicted information on the object's state. As expected, we obtain the usual prediction formula for random variables.
      
      We shall also suppose that the \ac{opm} $\bar{O}_k(\cdot\given X_k = x)$, representing the information the analyst possesses on the sensor's observation process, is based on a single possibility function $h_k(\cdot|x)$ on $\boY_k$. The update formula \eqref{eq:upd_opm} then reduces to
      \eqnl{eq:upd_red}
      {
        p_{k}(x \given y) = \dfrac{h_k(y \given x)p_{k\given k-1}(x)}{\int h_k(y \given x')p_{k\given k-1}(x')\d x'},
      }
      where $p_{k}$ is the \ac{pdf} describing the updated information on the object's state\footnote{The explicitly dependency of $p_{k}$ over past observations is omitted here for the sake of simplicity}. As expected, we obtain an update equation with a very similar structure than the usual Bayes' update rule, except that the observation process is described by a possibility function $h_t$ on $\boY_k$ rather than a likelihood function $\ell_t$, i.e., a \ac{pdf} on $\boY_k$.
      
  \section{Modeling of \ac{ssa} data sources with \acp{opm}} \label{sec:modeling}
    In this section, we shall focus on the modeling of two data sources relevant to the context of \ac{ssa}, namely, a radar ground station and \ac{usscom}'s catalog. Both data sources provide point-like observations in a well-defined observation space. They differ, however, on the available information about their observation process: a statistical description of the accuracy of a radar-like sensor is usually available, while no measure of uncertainty is provided (at least publicly) alongside \acp{tle}. We will see in that section that both data sources can be represented in a natural way with a single possibility function.
    
    \subsection{Radar observations}
      We shall simulate, in the scope of this paper, a radar ground station with Doppler effect. Suppose that a radar is exploited at some time $k$ relevant to the scenario. The observation space $\boY_{k} \subset \bbR^4$ describes a radar-like return $y = [\rho, \theta, \varphi, \dot{\rho}]$, providing range, azimuth, elevation, and range rate coordinates in the topocentric frame centered on the radar ground station on the surface of Earth.
      
      As usual for radar- or telescope-like sensors, the analyst is assumed to have access to some statistical description of the accuracy of the observation process, given by some $4\times 4$ positive-definite covariance matrix $S_t$. The possibility function $h_k(\cdot|x)$ on $\boY_{k}$ in Eq.~\eqref{eq:upd_red} is simply given by the Gaussian possibility
      \eqnla{eq:radar_poss}
      {
         &h^{\rad}_k(y|x) = \bar\calN(y; o_k(x), S_k)
         \\
         &= \exp\Big( -\dfrac{1}{2} (y-o_k(x))^{\tr}S_k^{-1}(y-o_k(x))\Big),
      }
      where $o_k$ is the mapping transforming coordinates from the target state space $\boX$ to the observation space $\boY_{k}$. Note that this possibility function does not \emph{characterize} the residual error of the radar observation process, a complex phenomenon that involves physical properties of the radar that may be known from the manufacturer, but also unknown perturbations due to the atmosphere, the object's shape, etc. Perhaps the radar observation is fully deterministic, or perhaps it possesses some aleatoric uncertainty that even an observer with perfect knowledge of the radar, the object, and the middle between them would still be unable to eliminate. In any case, a characterization of the radar's noise profile is unavailable to the analyst, who is only able to \emph{bound} the error in the observation process with \eqref{eq:radar_poss}.
      
      Note that the possibility \eqref{eq:radar_poss} is proportional to the likelihood
      \eqnl{eq:radar_likelihood}
      {
	\ell^{\rad}_k(y|x) = \dfrac{1}{\sqrt{|2\pi S_t|}}h^{\rad}_t(y|x),
      }
      i.e., the \emph{probabilistic} description usually adopted for a radar described by $S_k$, \emph{characterizing} the radar's noise profile. The difference is important: while the possibility $h_k$ is a \emph{dimensionless} function on $\boY_k$, the likelihood $\ell_k$ is a \ac{pdf} on $\boY_k$ and has dimension as the inverse of the unit volume of the observation space $\boY_k$. The value of the likelihood in Eq.~\eqref{eq:radar_likelihood} thus \emph{scales} with the changes on the reference measure on $\boY_t$ -- say, if the azimuth and elevation angles are measured in arc-seconds rather than in degrees -- while the possibility in Eq.~\eqref{eq:radar_poss} \emph{does not} \citep{Delande_E_2017_1_2}. This has important consequences in the context of multi-target detection and tracking problems, as briefly discussed in Sec.~\ref{sec:discussion}.
 
    \subsection{\acp{tle}}
      We shall query, in the scope of this paper, real \ac{tle} data produced by the \ac{usscom}\footnote{Available at \url{https://www.space-track.org}.}. For the sake of simplicity, we shall convert the original orbital elements of the \acp{tle} to Cartesian coordinates in the reference \ac{eci} frame, so that a \ac{tle} is considered as a point-like observation $y$ in the observation space $\boY_t = \boX$. Unlike radar-like observations, little is known about the generation process behind the \acp{tle}, let alone a statistical description of their residual error, i.e., the discrepancy between a \ac{rso}'s dynamical state $x$ and the generated \ac{tle} $y$.
      
      In order to produce some quantitative description of the \acp{tle}'s accuracy, we performed a brief analysis on the satellite 0E0E operated by Planet Labs\footnote{\url{https://www.planet.com/}.}, on the $4$-day-long period starting from Sep 2, 2017 at midnight. Using a batch least-square method\footnote{Details on the orbital propagator employed are given in the description of the tracking algorithm in Sec.~\ref{sec:scenario}.}, an orbital trajectory for that satellite was produced from \ac{gps} points provided by Planet Labs, considered as ground truth for the analysis to come. We then queried the \ac{usscom} for the $17$ \ac{tle} points available relating to that satellite over that period, and studied the discrepancies of the \acp{tle} as explained in the following paragraphs.

      The natural observation space of a \ac{tle} is six-dimensional, since it provides information on the full dynamical state of the observed \ac{rso}. In this paper, we shall propose a simple possibility function $h^{\tle}$ of the observation process that aggregates the information extracted from a \ac{tle} into two meaningful dimensions, capturing salient features of the \ac{rso}'s orbital regime. Both the \emph{specific angular momentum} $\bsh$ and the \emph{specific orbital energy} $\epsilon$ are conservative quantities of an orbital trajectory, in the two-body problem and assuming Keplerian motion. Denoting by $\bsp$ (\ac{resp} $\bsv$) the vector of position (\ac{resp} velocity) coordinates of the \ac{rso} with state $x = [\bsp, \bsv]$, $\bsh$ and $\epsilon$ are given by
      \eqnla{eq:specific_q}
      {
	\bsh(x) &= \bsp \times \bsv,
	\\
	\epsilon(x) &= -\dfrac{\mu_{\e}}{||\bsp||} + \dfrac{||\bsv||^2}{2}, \label{eq:specific_q_energy}
      }
      where $\mu_{\e}$ denotes Earth's gravitational constant.
      
\begin{figure}[ht]
	\centering
	\includegraphics[width=\imagewidth]{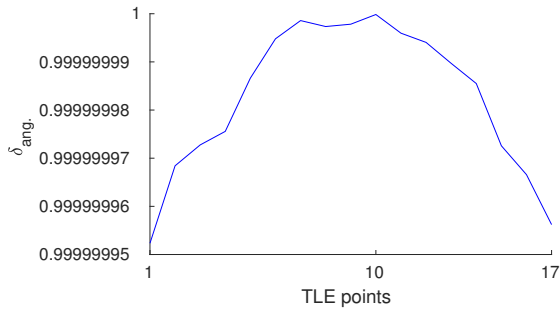}
	\caption{Angle offset $\delta_{\ang}$ of the \acp{tle} \ac{wrt} to the ground truth state, for Planet Labs' 0E0E satellite and over a $4$-day-long period.\label{fig:tle_angle}}	
      \end{figure}

      Because \acp{tle} are expected to provide accurate information on the \ac{rso}'s orbital plane, the first component on which we shall build the possibility function $h^{\tle}$ is the normalized dot product
      \eqnl{eq:delta_h}
      {
	\delta_{\ang}(y|x) = \dfrac{\bsh(y)}{||\bsh(y)||} \cdot \dfrac{\bsh(x)}{||\bsh(x)||},
      }
      quantifying the angle offset between the orbital planes described by some \ac{tle} $y$ and some \ac{rso}'s state $x$. The study of the angle offset \eqref{eq:delta_h} on the training data is depicted in Fig.~\ref{fig:tle_angle}.
      
\begin{figure}[ht]
	\centering
	\includegraphics[width=\imagewidth]{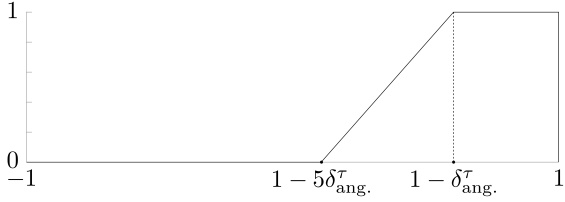}
	\caption{A possibility function $h^{\tle}_{\ang}$ on the interval $[-1,1]$ representing the information on the orbital angle provided by a \ac{tle} point. The tolerance bound $\delta^{\tau}_{\ang}$ is selected from the $4$-day-long training period with Planet Labs' 0E0E satellite.\label{fig:poss_angle}}	
      \end{figure}

      The analysis of the angle offset \eqref{eq:delta_h} on the training period (Fig.~\ref{fig:tle_angle}) suggests that the \acp{tle} do indeed provide accurate information on the \ac{rso}'s orbital plane, with an extremely small deviation of magnitude $\delta^{\tau}_{\ang} = 10^{-7}$ below the nominal (and maximal) value of $\delta^{\n}_{\ang} = 1$. We can then represent this information with a possibility function $h^{\tle}_{\ang}$ on the interval $[-1,1]$ as shown in Fig.~\ref{fig:poss_angle}.
      
      The possibility function in Fig.~\ref{fig:poss_angle} reflects our acquired information on the observation process; since it is limited in scope, we follow a cautious approach and model a possibility that has the maximum value of $1$ around the values consistent with the training data, but decreases linearly towards zero over a larger region that covers five times the deviation $\delta^{\tau}_{\ang}$ observed on the training data.
      
      Similarly to the radar model, this possibility function does not \emph{characterize} the residual error in the \ac{rso}'s orbital plane as given by a \ac{tle} point, like a typical likelihood function would. Such a descriptive model remains largely inaccessible through the modest analysis in Fig.~\ref{fig:tle_angle}, limited to a small period of time only and to a single satellite only. Whether a statistical profile of a \ac{tle}'s residual error could be inferred from a large scale analysis is an interesting question beyond the scope of this paper.
      
\begin{figure}[ht]
	\centering
	\includegraphics[width=\imagewidth]{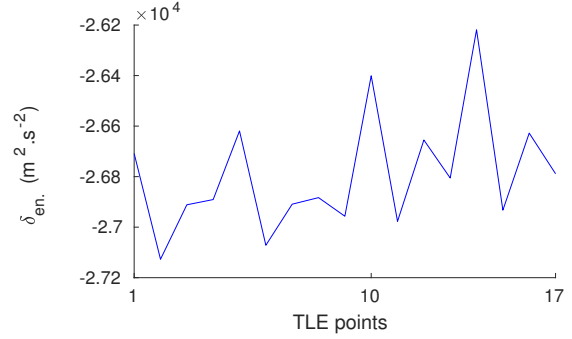}
	\caption{Energy offset $\delta_{\en}$ of the \acp{tle} \ac{wrt} to the ground truth state, for Planet Labs' 0E0E satellite and over a $4$-day-long period.\label{fig:tle_energy}}	
      \end{figure}

      A similar analysis can be performed on the specific orbital energy in Eq.~\eqref{eq:specific_q_energy}. The second component on which we shall build the possibility function $h^{\tle}_{\en}$ is the energy offset
      \eqnsml{eq:delta_energy}
      {
	\delta_{\en}(y|x) = -\mu_{\e}\left(||\bsp(y)||^{-1} - ||\bsp(x)||^{-1}\right)
	\\
	+ \dfrac{1}{2}\left(||\bsv(y)||^2 - ||\bsv(x)||^2\right),
      }
      which represents the difference in specific orbital energies between the orbits described by some \ac{tle} $y$ and some \ac{rso}'s state $x$. The study of the energy offset \eqref{eq:delta_energy} on the training data yields the results in Fig.~\ref{fig:tle_energy}.
      
\begin{figure}[ht]
	\centering
	\includegraphics[width=\imagewidth]{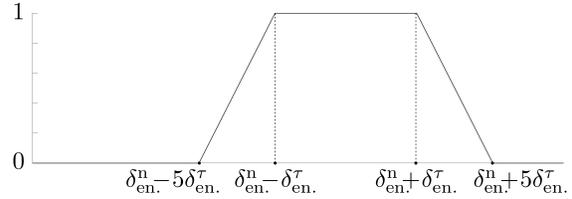}
	\caption{A possiblity function $h^{\tle}_{\en}$ on $\bbR$ representing the information on the specific orbital energy provided by a \ac{tle} point. The nominal value $\delta^{\n}_{\en}$ and the tolerance bound $\delta^{\tau}_{\en}$ are selected from the $4$-day-long training period with Planet Labs' 0E0E satellite.\label{fig:poss_energy}}	
      \end{figure}

      The energy offset on the training period suggests an underestimation from the collected \acp{tle} \ac{wrt} to the (supposed) ground truth, that amounts to some $2.67\times 10^4\si{\meter\squared\per\second\squared}$, with a deviation of about $0.5\times 10^4\si{\meter\squared\per\second\squared}$. Similarly to the study of the angle offset, we can represent this information with a possibility function $h^{\tle}_{\en}$ on $\bbR$ as shown in Fig.~\ref{fig:poss_energy}.
      
      Interpreting the information produced by a \ac{tle} solely through the orbital plane and the specific orbital energy of the observed \ac{rso}, we can then design a possibility function $h^{\tle}$ for a ``\ac{tle}-generator'' as
      \eqnl{eq:poss_tle_1}
      {
	h^{\tle}(y|x) = h^{\tle}_{\ang}(\delta_{\ang}(y|x))h^{\tle}_{\en}(\delta_{\en}(y|x)).
      }
      It is easy to see that $h^{\tle}(\cdot|x)$ in Eq.~\eqref{eq:poss_tle_1} is indeed a possibility function; in particular, it reaches the maximum value of $1$ on the \ac{tle} space $\boY$.
      
  \section{Scenario implementation} \label{sec:scenario}
    This section details the implementation of the Bayesian estimation scenario on which the fusion of \acp{tle} and radar-like observations will be exploited for the detection and tracking of a single orbiting \ac{rso}.
    
    \subsection{Ground truth and observations}
      As for the training data on which the analysis of the \acp{tle} was performed in Sec.~\ref{sec:modeling}, the reference \ac{rso} is Planet Labs' 0E0E satellite (the corresponding CAT ID assigned by the \ac{usscom} is 41609). The filtering period covers approximately one day, from Sep 6, 2017 00:11:47 (UTC) to Sep 7, 2017 00:00:00 (UTC). A ground truth trajectory was produced for that period from a batch least-square method, exploiting the \ac{gps} points provided by Planet Labs.
      
      The $5$ \ac{tle} points relevant to the filtering period were queried from the \ac{usscom}; their state are given in Tab.~\ref{tab:ground_truth_data}. A radar with Doppler effect is simulated in order to generate radar observations, following the characteristics of LeoLabs' PFISR given in \cite{Nicolls_M_2017_1}. Its \ac{fov} is set as the ball centered on Fairbanks (Alaska, U.S.A.) with a radius of $\SI{2000}{\kilo\meter}$, and the \ac{rso} is observed during eight brief windows over the filtering period (see Fig.~\ref{fig:map_error}). The noise covariance matrix of the radar is set as the diagonal matrix $S_t = \mathrm{diag}([\sigma^2_{\rho}, \sigma^2_{\theta}, \sigma^2_{\varphi}, \sigma^2_{\dot{\rho}}])$, where the standard deviations are set to the values $\sigma_{\rho} = \SI{28}{\meter}$, $\sigma_{\theta} = \SI{0.1}{\degree}$, $\sigma_{\varphi} = \SI{0.1}{\degree}$, $\sigma_{\dot{\rho}} = \SI{11}{\meter\per\second}$.
      
      \begin{table}[ht]
	\caption{\acp{tle} points for Planet Labs' OEOE satellite, queried from the \ac{usscom}'s catalog (CAT ID 41609).\label{tab:ground_truth_data}}
	\centering
	{
	  \footnotesize
	  \begin{tabular}{cccccc}
	    $\Omega~(\si{\degree})$ & $i~(\si{\degree})$ & $\omega~(\si{\degree})$ & $n~(\si{\radian\per\sec})$ & $e$ & $M~(\si{\degree})$
	    \\
	    \hline
	    \multicolumn{6}{c}{\textcolor{Gray}{1st \ac{tle} (2017-09-06 03:58:22)}}
	    \\
	    311.18 & 97.45 & 144.12 & 11.07e-4 & 11.95e-4 & 216.09
	    \\
	    \hline
	    \multicolumn{6}{c}{\textcolor{Gray}{2nd \ac{tle} (2017-09-06 11:51:33)}}
	    \\
	    311.51 & 97.45 & 143.20 & 11.07e-4 & 11.96e-4 & 217.00
	    \\
	    \hline
	    \multicolumn{6}{c}{\textcolor{Gray}{3rd \ac{tle} (2017-09-06 13:26:11)}}
	    \\
	    311.57 & 97.45 & 143.01 & 11.07e-4 & 11.97e-4 & 217.20
	    \\
	    \hline
	    \multicolumn{6}{c}{\textcolor{Gray}{4th \ac{tle} (2017-09-06 18:10:06)}}
	    \\
	    311.77 & 97.45 & 142.49 & 11.07e-4 & 11.99e-4 & 217.72
	    \\
	    \hline
	    \multicolumn{6}{c}{\textcolor{Gray}{5th \ac{tle} (2017-09-06 21:19:22)}}
	    \\
	    311.90 & 97.45 & 141.88 & 11.07e-4 & 12.01e-4 & 218.33
	    \\
	  \end{tabular}
	}
      \end{table}
    
    \subsection{Filter implementation}
      The object state space $\boX \subset \bbR^6$ describes the \ac{rso}'s position and velocity coordinates in the reference \ac{eci} frame. The time flow of the filtering scenario is built as follows: the filtering period is split in even time lapses of $\SI{120}{\second}$, to which the five collection dates of \acp{tle} are added. The resulting time flow is indexed with $k \in \bbN$ with $k = 0$ corresponding to the scenario's initial date. In addition, the duration since epoch J2000 (in $\si{\second}$) is given at time $k$ by $t_k$.
      
      \subsubsection{Time prediction step}
	Since the law $p_k$ on $\boX$ describing the \ac{rso}'s state is not parameterizable in a straightforward manner, the practical implementation of the Bayesian filtering equations follows a \ac{smc} approach. At time $k-1$, the posterior \ac{pdf} $p_{k-1}$ is approximated by a set of $N = 500$ weighted particles $\{w^{(i)}_{k-1}, x^{(i)}_{k-1}\}_{i = 1}^N$ such that
	\eqns
	{
	  p_{k-1}(\cdot) \simeq \sum_{i = 1}^N w^{(i)}_{k-1} \delta_{x^{(i)}_{k-1}}(\cdot),
	}
	where $\delta_x$ is the Dirac function at $x \in \boX$, and with $\sum_{i = 1}^N w^{(i)}_{k-1} = 1$.
	
	The prediction kernel $m_k$ in Eq.~\eqref{eq:pred_red} aims at describing a \ac{leo} trajectory between epochs $t_{k-1}$ and $t_k$, and is constructed as follows. Assuming the object has state $x = \left[\bsp(t), \bsv(t)\right]$ at some epoch $t$, the acceleration vector $\dot{\bsv}(t)$ is given by
	\eqns
	{
	  \dot{\bsv}(t) = \bsa_{\mod}(\bsp(t), \bsv(t), t),
	}
	where the mapping $\bsa_{\mod}$ computes the orbital acceleration terms modeled in the scope of this paper. In addition to the central term of Earth's gravitational pull, it includes the following perturbations: the zonal/tesseral effects up to order and degree $20$, the gravitational pull of the Sun and the Moon, the solar radiation pressure (assuming an \ac{amr} of $\SI{0.0015}{\meter\squared\per\kilo\gram}$, and a spherical shape with a radiation pressure coefficient of $0.3$), and a drag term based on the MSIS86 atmospheric model (assuming a ballistic coefficient of $\SI{25}{\kilo\gram\per\meter\squared}$). An additional acceleration term, accounting for the unmodeled perturbations, is built as
	\eqns
	{
	  \bsa_{\epsilon}(\bsp(t), \bsv(t), t, \bsomega(t_{k-1})) = f_{\ric(\bsp, \bsv)}^{\eci}\left((t - t_{k-1})\bsomega(t_{k-1})\right),
	}
	where $\bsomega$ is a zero-mean Gaussian noise with standard deviation $\sigma_{\r} = \sigma_{\i} = \sigma_{\c} = \SI{e-5}{\metre\per\second\cubed}$ on each component in the object's \ac{ric} frame, and $f_{\ric(\bsp, \bsv)}^{\eci}$ is the mapping that transforms a vector in the object's \ac{ric} frame to the reference \ac{eci} frame. Denoting by
	\eqnsm
	{
	  \bsa_{\orb}\left(\bsp(t), \bsv(t), t, \bsomega(t_{k-1})\right)
	  \\
	  = \bsa_{\mod}\left(\bsp(t), \bsv(t), t) + \bsa_{\epsilon}(\bsp(t), \bsv(t), t, \bsomega(t_{k-1})\right)
	}
	the total acceleration term describing the orbital dynamics, the time-derivative $\dot{x}$ of the target state is then given by
	\eqns
	{
	  [\dot{\bsp}(t), \dot{\bsv}(t)] = \left[\bsv(t), \bsa_{\orb}\left(\bsp(t), \bsv(t), t, \bsomega(t_{k-1})\right)\right].
	}
	Given a particle $x^{(i)}_{k-1} = [\bsp^{(i)}(t_{k-1}),\bsv^{(i)}(t_{k-1})]$, the predicted particle $x^{(i)}_{k|k-1}$ is then computed through
	\eqnsml{eq:pred_part}
	{
	  x^{(i)}_{k|k-1} = [\bsp^{(i)}(t_{k-1}),\bsv^{(i)}(t_{k-1})]
	  \\
	  +\int_{t_{k-1}}^{t_k} \left[\bsv^{(i)}(t), \bsa_{\orb}\left(\bsp^{(i)}(t), \bsv^{(i)}(t), t, \bsomega^{(i)}(t_{k-1})\right) \right] \d t.
	}
	Since the stochastic component $\bsomega^{(i)}(t_{k-1})$ is constant throughout the time interval $[t_{k-1}, t_k]$, Eq.~\eqref{eq:pred_part} can be solved with a usual numerical integrator (we used Matlab\textsuperscript{\tiny\textregistered}'s ode45).
	
      \subsubsection{Data update step}
	At time $k-1$, the predicted \ac{pdf} $p_{k|k-1}$ is thus approximated by the set $\{w^{(i)}_{k-1}, x^{(i)}_{k|k-1}\}_{i = 1}^N$. Three cases are now to be considered, depending on the availability of corrective data:
	
	\underline{Case 1. No observation is available}\newline
	This is by far the most frequent case, and also the most straightforward to process. Since no additional information is available on the \ac{rso} the prior $p_k$ is set as the posterior $p_{k|k-1}$, i.e.
	\eqns
	{
	  x^{(i)}_{k} = x^{(i)}_{k|k-1},\quad w^{(i)}_{k} = w^{(i)}_{k-1},
	}
	for any $1 \leq i \leq N$.
	  
	\underline{Case 2. A \ac{tle} $y$ is available}\newline
	This is also straightforward to process. The update equation \eqref{eq:upd_red} becomes
	\eqns
	{
	  x^{(i)}_{k} = x^{(i)}_{k|k-1},\quad w^{(i)}_{k} = \dfrac{h^{\tle}(y \given x^{(i)}_{k|k-1})w^{(i)}_{k-1}}{\sum_{j=1}^N h^{\tle}(y \given x^{(i)}_{k|k-1})w^{(i)}_{k-1}},
	}
	for any $1 \leq i \leq N$.
	
	\underline{Case 3. A radar observation $y$ is available}\newline
	The data update equation in this case is similar to the \ac{tle}'s above, except that the radar possibility $h^{\rad}_k$ is substituted to the \ac{tle} possibility $h^{\tle}$. However, because of the high accuracy of radar observations, implementing the data update mechanism leads to quick degeneracy among the particles, especially when the particle cloud has spread significantly after a long period without observations \citep{Delande_E_2017_1_2}. Our approach is to find a parametrization of the predicted \ac{pdf} $p_{k|k-1}$ leading to a more robust data update mechanism unaffected by the scarcity of particles.
	
	Since \acp{pdf} representing orbital states can hardly be parametrized in a simple manner in the Cartesian coordinates $\boX$, we exploit an alternative space $\boS_k$ -- namely, the spherical coordinates in the sensor's local frame -- in which a Gaussian approximation is more valid. The data update procedure can be summarized as follows \citep{Delande_E_2017_1_2}:
	\begin{enumerate}[nolistsep]
	 \item Transform $p_{k|k-1}$ in the spherical frame $\boS_k$: $\{w^{(i)}_{k|k-1}, x^{(i)}_{k|k-1}\}_{i=1}^N \rightarrow \{w^{(i)}_{k|k-1}, s^{(i)}_{k|k-1}\}_{i=1}^N$,
	 \item Approximate as a Gaussian distribution: $\{w^{(i)}_{k|k-1}, s^{(i)}_{k|k-1}\}_{i=1}^N \rightarrow (\mu_{k|k-1}, Q_{k|k-1})$,
	 \item Update Gaussian distribution, using observation $y$: $(\mu_{k|k-1}, Q_{k|k-1}) \rightarrow (\mu_k, Q_k)$,
	 \item Sample resulting distribution: $(\mu_k, Q_k) \rightarrow \{N^{-1}, s^{(i)}_{k}\}_{i=1}^N$,
	 \item Transform back in the object space $\boX$: $\{N^{-1}, s^{(i)}_{k}\}_{i=1}^N \rightarrow \{N^{-1}, x^{(i)}_{k}\}_{i=1}^N$.
	\end{enumerate}
	Note that the data update step in the spherical frame $\boS_k$ is a simple linear Kalman update, since the radar is modeled with a Gaussian possibility in Eq.~\eqref{eq:radar_poss}.
		
	In any case, the particles are resampled, following the data correction step, if the efficiency ratio $r_{\eff} = \frac{1}{N\sum_{i = 1}^{N}(w^{(i)}_{k})^{2}}$ falls below a threshold set at $20\%$.
	
      \subsubsection{Initial orbit determination}
	No prior information is assumed on the \ac{rso}'s state prior to its first observation, and an initial orbit determination procedure is implemented in order to determine the posterior distribution following the first radar observation. We followed the admissible region approach developed in \cite{DeMars_KJ_2013_2}, where admissible values for the unobserved angular rates $\dot{\theta}, \dot{\varphi}$ are established from the initial radar observation $y = [\rho, \theta, \varphi, \dot{\rho}]$, exploiting internal energy constraints \citep{Tommei_G_2007, Farnocchia_D_2010}. We employed a similar \ac{smc} implementation of the admissible region approach as in our previous works; more details can be found in \cite{Delande_E_2017_1_2}.
	
  \section{Filtering results} \label{sec:simulation}
    In order to illustrate the exploitation of \acp{tle} as a complementary data source, we ran the Bayesian tracking filter described in Sec.~\ref{sec:scenario} on two parallel scenarios based on the same ground truth trajectory: one exploits the simulated radar observations only, the other exploits the radar observations \emph{and} the \ac{tle} points collected from the \ac{usscom}'s catalog. The filters' outputs are averaged over \nosmcrun runs, where the observations generated for each run follow the radar model presented in Sec.~\ref{sec:scenario}.
    
\begin{figure}[ht]
      \centering
      \includegraphics[width=\imagewidth]{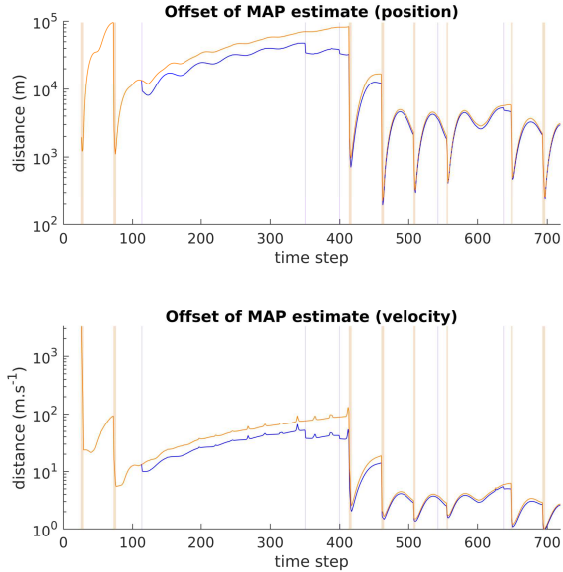}
      \caption{Distance between the \ac{map} estimate of the filter's output and the ground truth. The blue (\ac{resp} red) vertical bars correspond to the periods where \acp{tle} (\ac{resp} radar observations) are available. The orange (\ac{resp} blue) plot corresponds to the filter using the radar only (\ac{resp} using the radar and the \ac{usscom}'s catalog).\label{fig:map_error}}	
    \end{figure}

    We first compare the output of the filter through a \ac{map} estimate of the \ac{rso}'s state. Because of the frequent resampling and the scarcity of data corrective steps, the posterior particle weights are often uniformly distributed; added to the fact that the statistical moments of a \ac{pdf} on $\boX$ have little sense in the context of \ac{ssa}\footnote{For example, even if the particle cloud spreads alongside a ray of close orbital trajectories, the weighted mean of the particle distribution may belong to an entirely different orbit.} it is not straightforward to derive a \ac{map} estimate directly from the posterior \ac{pdf} $p_k$. Similarly to the processing of radar observations presented in Sec.~\ref{sec:scenario}, we use the expression of the posterior \ac{pdf} in spherical coordinates (still in the reference \ac{eci} frame), assumed Gaussian distributed, to derive a \ac{map} estimate of the \ac{rso}'s state. The procedure can be summarized as follows:
    \begin{enumerate}[nolistsep]
      \item Transform $p_{k}$ to spherical coordinates: $\{w^{(i)}_k, x^{(i)}_k\}_{i=1}^N \rightarrow \{w^{(i)}_k, s^{(i)}_k\}_{i=1}^N$,
      \item Compute the \ac{map} in spherical coordinates as the weighted mean of the resulting distribution: $\{w^{(i)}_k, s^{(i)}_k\}_{i=1}^N \rightarrow s_\map = \sum w^{(i)}_k s^{(i)}_k$,
      \item Transform back the \ac{map} estimate to Cartesian coordinates: $s_\map \rightarrow x_\map$.
    \end{enumerate}
    Note that the resulting point $x_\map \in \boX$ is not, strictly speaking, the \ac{map} estimate of the \ac{rso}'s state, but will be considered as such for the purpose as this analysis. The \ac{map} estimate of the filter's output is depicted in Fig.~\ref{fig:map_error}. 
    
\begin{figure}[ht]
      \centering
      \includegraphics[width=\imagewidth]{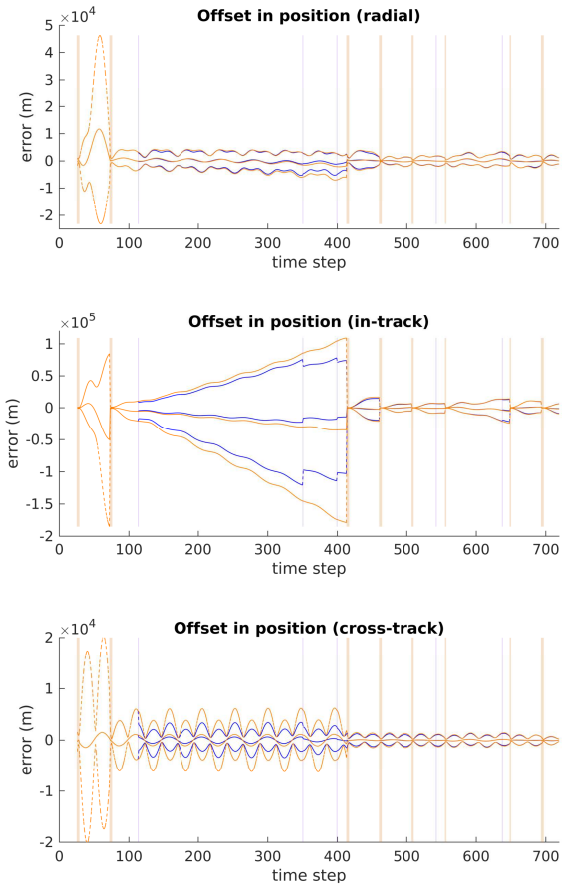}
      \caption{Offset (mean +/- std. dev.) of the filter's output in the ground truth's \ac{ric} frame (position coordinates). The blue (\ac{resp} red) vertical bars correspond to the periods where \acp{tle} (\ac{resp} radar observations) are available. The orange (\ac{resp} blue) plot corresponds to the filter using the radar only (\ac{resp} using the radar and the \ac{usscom}'s catalog).\label{fig:ric_pos_error}}	
    \end{figure}

\begin{figure}[ht]
      \centering
      \includegraphics[width=\imagewidth]{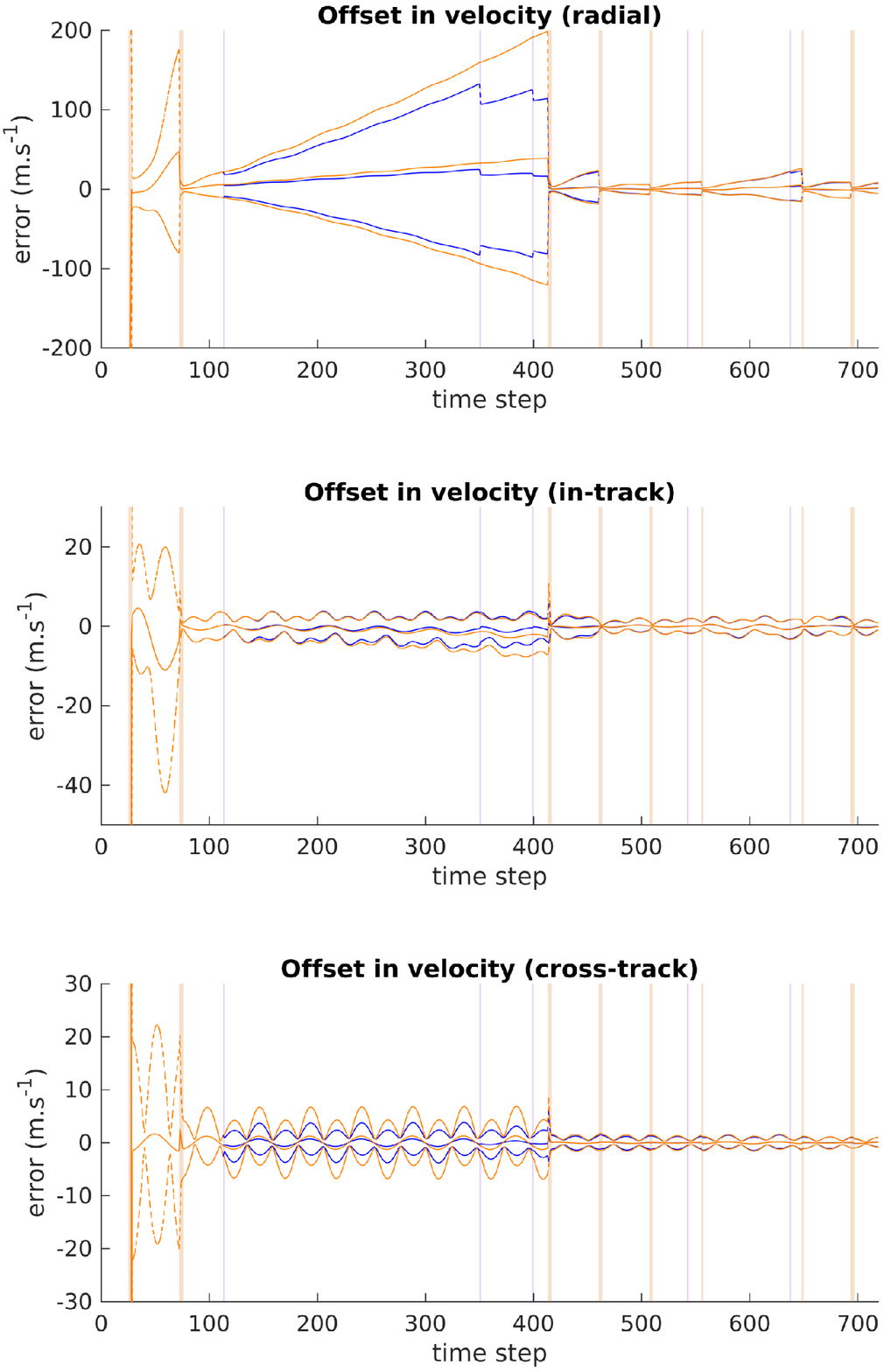}
      \caption{Offset (mean +/- std. dev.) of the filter's output in the ground truth's \ac{ric} frame (velocity coordinates). The blue (\ac{resp} red) vertical bars correspond to the periods where \acp{tle} (\ac{resp} radar observations) are available. The orange (\ac{resp} blue) plot corresponds to the filter using the radar only (\ac{resp} using the radar and the \ac{usscom}'s catalog). The plots are cut on the y-axis during the first pass of the \ac{rso} in the sensor \ac{fov}.\label{fig:ric_vel_error}}
    \end{figure}

    We also compare the output of the filter in the \ac{ric} frame of the ground truth trajectory; the result is depicted in Fig.~\ref{fig:ric_pos_error} for position coordinates, and in Fig.~\ref{fig:ric_vel_error} for velocity coordinates.
    
    By construction, the output of the two filters are nearly identical, from the first pass of the \ac{rso} in the radar \ac{fov} to the first data correction with a \ac{tle} point. Since the radar does not provide information on the \ac{rso}'s angular rate, the initial estimation on the \ac{rso}'s state is poor and the associated uncertainty grows significantly as soon as the \ac{rso} leaves the radar \ac{fov} and observations are unavailable. As expected, the following data correction steps with a \ac{tle} improve the quality of the \ac{map} estimate (see Fig.~\ref{fig:map_error}), and sharpen the distribution significantly (see Figs~\ref{fig:ric_pos_error}, \ref{fig:ric_vel_error}).
    
    Due to the cautious approach in the modeling of the \acp{tle} (see Sec.~\ref{sec:modeling}), the information gain of the following \ac{tle} points is much less significant, though not inexistent, once the \ac{rso} has entered the radar \ac{fov} for the third time (from around time step $400$, onwards). This suggests that a more informative \ac{tle} model than presented in Sec.~\ref{sec:modeling}, built on a larger set of training data and/or on different physical quantities, may benefit the estimation to a larger extent.
  
  \section{Future works} \label{sec:discussion}
    The introduction of \acp{opm} to the context of \ac{ssa} opens a wealth of possibilities for the development of Bayesian estimation algorithms, as the modeling of uncertain components in a typical \ac{ssa} target tracking scenario can be revisited through \acp{opm} (and the associated possibility functions) rather than \acp{pdf}. 
    
    Many differences between the two approaches are apparent in multi-target detection and tracking scenarios, where marginalized quantities play an important role in the computation of the relative weight (i.e., probability of existence) of individual tracks. For example, because possibility functions are \emph{dimensionless}, a marginalization term such as the denominator $\int h^{\rad}_t(y|x) p_{t|t-1}(x)\d x$ in Eq.~\eqref{eq:upd_red}, assessing the matching between the estimated object and the observation $y$, does not scale whether the distances in the radar observation space are measured in meters or kilometers, the angles in degrees or radians, etc. The usual denominator $\int \ell^{\rad}_t(y|x) p_{t|t-1}(x)\d x$ does, however, since the \ac{pdf} $\ell^{\rad}_t(y|x)$ in Eq.~\eqref{eq:radar_likelihood} has units and thus scales with the change of reference measure on the observation space. This feature is often overlooked, but raises issues when such a data association event must be compared to another uncertain event that either scales differently (e.g., the association of the same estimated object to a telescope-like observation $y'$), or does not scale at all (e.g., a missed detection event) with arbitrary changes in physical units \citep{Delande_E_2017_1_2}. 
    
    A natural follow-up on our current work would be to refine the modeling of the \acp{tle} as an uncertain data source. The analysis presented in this paper could be extended to larger period of time and to a wider range of \acp{rso} covering different orbital regimes; studies on the accuracy of \acp{tle} such as in \cite{Kelso_TS_2007_1, Fruh_C_2012_2} would also provide ground for a further refinement of the \ac{tle} model. Another natural follow-up would be to propose an \ac{opm}-based representation of other data sources already exploited in the context of \ac{ssa}, such as telescopes, cameras, etc.  Less conventional data sources can be found in the numerous natural language statements related to specific launch events, the ill-formatted data collected from unknown/untrusted amateur telescopes, etc. Natural language statements have already been approached with \acp{opm} in the design of target tracking algorithms in \cite{Bishop_AN_2018_1}, and this study could spearhead the integration of a range of data sources relevant to the context of \ac{ssa}.
    
    Another promising lead to follow is to explore an \ac{opm}-based representation of the propagated information, currently represented with a usual \ac{pdf} $p_k$ in Eqs~\eqref{eq:pred_red}, \eqref{eq:upd_red}. The initial orbit determination procedure \citep{DeMars_KJ_2013_2} leaves no real knowledge of the distribution of the \ac{rso}'s state \emph{within} the admissible region: this is a straightforward example where the information possessed by the analyst is represented in a more natural and less prejudiced manner with a possibility function equal to one within the admissible region and zero outside, rather than a uniform \ac{pdf} supported by the admissible region.
    
    Orbital propagators designed for practical \ac{ssa} tracking algorithms always operate on limited information. They approximate physical effects affecting the orbital trajectory (e.g. limited order in the zonal/tesseral components of the Earth's gravitational pull), discard them on purpose (e.g. the Earth's shadow is ignored) or, less conspicuously, ignore perturbation effects whose existence is simply unknown to the space analyst. Given the complexity of the orbital dynamics and the many sources of uncertainty shaping the design of an orbital propagator, a probabilistic representation of the prediction model appears over-descriptive and could benefit from an \ac{opm}-representation as well. 
  
  \section{Conclusion} \label{sec:conclusion}
    In this paper, we introduce a new representation of uncertainty for Bayesian detection and tracking algorithms in the context of \ac{ssa}, based on \emph{\aclp{opm}} rather than \emph{\aclp{pdf}}. Less descriptive than \acp{pdf}, \acp{opm} do not characterize the distribution of an uncertain system's state; rather, they aim at proposing a prejudice-free representation of the system's state, matching the information possessed by some operator -- say, a space analyst -- about that system and \emph{nothing more}. We show that each uncertain component of a single object Bayesian estimation problem -- the estimated state, the prediction model, the data source(s) generating observations/opinions -- can be represented either with an \ac{opm} or a \ac{pdf} in a single coherent Bayesian estimation framework.
    
    More versatile than \acp{pdf}, \acp{opm} allow for the modeling of uncertain components for which statistical information is scarce or inexistent. In particular, we show that \acp{tle} can be treated as a data source in a Bayesian estimation problem through a simple \ac{opm}-based model, and fused with radar observations in a \ac{rso} tracking algorithm. This concept is then illustrated on a scenario where \ac{tle} points queried from the \ac{usscom} catalog and simulated radar observations are exploited in order to track a \ac{leo} satellite.
  
  \section*{Acknowledgements}
    The authors wish to thank Vivek Vittaldev (Planet Labs) for his help in accessing and intepreting \ac{gps} data collected by Planet Labs, and Marcus Bever (University of Texas at Austin) for the extraction of \ac{usscom}'s \acp{tle} and Planet Labs' \ac{gps} points, and their formatting into a common reference frame.

  \section*{References}
    \bibliographystyle{elsarticle-harv.bst}
    \bibliography{bibliography.bib}
\end{document}